# Uncertainty Analysis for Material Measurements Using the Vector Network Analyzer

Nosherwan Shoaib[*#], Marco Sellone[*], Luciano Brunetti[*] and Luca Oberto[*]

[*]Nanoscience and Materials Division, National Institute of Metrological Research - INRiM, Strada delle cacce 91, Torino, ITALY
E-mail: m.sellone@inrim.it, l.brunetti@inrim.it and l.oberto@inrim.it

[#]Petroleum Institute, P. O. Box 2533, Abu-Dhabi, UAE
Email: nshoaib@pi.ac.ae

Abstract

This paper presents the characterization measurements and related uncertainty evaluation of a non-magnetic material using the Vector Network Analyzer (VNA) at microwave frequencies. The permittivity of the material under test is computed from the scattering parameters (S-parameters). The aim of the work is to highlight the different uncertainty contributions affecting the permittivity.

## 1. Introduction

The measurements of dielectric properties of materials at radio frequency have gained importance in the RF & microwave research field. The dielectric measurements are useful because they provide important information about the electric and magnetic properties of the materials under test (MUTs). The VNAs are widely used for material measurements and provide the material characterization in terms of scattering parameter (S-parameters) [1]-[4]. The S-parameters are further post-processed to obtain the permittivity of the MUTs [4]. However, the complete uncertainty estimation for permittivity measurements using the VNAs is still an open and active research topic. Even the recent works on material measurements using the VNA system are limited to the uncertainty due to VNA residual deviations of the pre-calibration only [3]. Therefore, the complete uncertainty budget showing different uncertainty contributions is still missing.

In this work we present a complete uncertainty evaluation along with uncertainty budgets for permittivity measurements of a non-magnetic material following a differential numerical approach [5]-[7]. The uncertainty analysis is carried out for complex-valued S-parameters. The S-parameter uncertainty is linearly propagated to the complex-valued permittivity, $\varepsilon_r = \varepsilon_r' - j\varepsilon_r''$, where $\varepsilon_r'$ and $\varepsilon_r''$ are the real and imaginary parts of $\varepsilon_r$ respectively. The $\varepsilon_r'$ is also known as dielectric constant, while the $\varepsilon_r''$ is usually referred to as the loss factor. Moreover, the dielectric loss tangent of the material sample (tan δ) can be written as:

$$\tan \delta = \frac{\varepsilon_r''}{\varepsilon_r'} \qquad (1)$$

The schematic of the measurement setup is shown in Fig. 1. A waveguide fixture is filled with the material sample and the S-parameters are measured. The layout of the waveguide fixture and the MUT is shown in Fig. 2.

**Fig. 1.** Schematic of the measurement setup

**Fig. 2.** Layout of the waveguide fixture and the MUT

The dimensional measurements of both the sample and the fixture are also performed. The propagation of uncertainties due to S-parameters and the dimensional measurements are carried out to compute the total uncertainty of the permittivity results. The uncertainty propagation is compliant with the "Guide to the expression of the uncertainty in measurements - (GUM)" [8]. The measurements are traceable to the SI (International System of units) through dimensional measurements. The traceability of the measurements is important for the comparison of the results obtained using different measurement systems. As an example, a sample of Teflon was used as MUT and a WR-90 VNA waveguide setup was used as measurement system. The complete uncertainty budgets at different frequencies are also presented for permittivity results.

## 2. Measurement Model and Uncertainty Analysis

The VNA measurement model used is described in [5] (see Eqn. 1) in which uncertainties are associated to the error terms. It includes the VNA noise, linearity, drift, cable stability and connector repeatability contributions related to S-parameters. This model has been implemented in a data acquisition software package. The measured transmission S-parameter ($S_{21}$) of the MUT in the X-band is further post-processed to compute the complex-valued permittivity in an iterative way by solving the following equation [4]:

$$S_{21} = e^{-\gamma_0(L_f - L)} \frac{T(1-\Gamma^2)}{1-\Gamma^2 T^2}, \quad (2)$$

where, $\Gamma = \frac{\gamma_0 - \gamma_1}{\gamma_0 + \gamma_1}$, $T = e^{-\gamma_1 L}$, $\gamma_0$ and $\gamma_1$ are the propagation constants in air and the material sample of length L, respectively. $L_f$ is the length of the waveguide fixture. $\gamma_0$ and $\gamma_1$ depend on the width of the waveguide fixture. The air gap corrections are also applied for the air gap that exists between the sample and the waveguide fixture. The air gap corrections are the function of the height of the waveguide fixture and the sample [4]. The dimensional measurements of the fixture and the sample have been performed and shown in Table 1. Their tolerance is 10 µm.

**Tab. 1.** Dimensional measurements of waveguide fixture and the Teflon sample (in mm).

The VNA setup has been previously characterized in terms of noise, linearity, drift, cable stability and connector repeatability and the different S-parameter uncertainty influences have been computed [9]. The permittivity results also include the uncertainty contribution due to dimensional tolerances of the fixture and the sample dimensions.

The uncertainty propagation from the S-parameters to the permittivity results has been carried out by using a general purpose library, known as Metas.UncLib [6] and a custom MATLAB [10] code. This library allows the propagation of variances of the input probability density functions (pdfs) through a measurement model, taking the correlations between the quantities fully into account. Metas.UncLib uses differentiation techniques and keeps track of the dependencies throughout the measurement model [6], [7].

## 3. Experimental Results

Measurements have been performed with a two-port VNA in a temperature and humidity controlled shielded room. The temperature and the relative humidity during the measurements were of $(23.0 \pm 0.3)°C$ and $(45 \pm 5)\%$ respectively. The VNA was calibrated using the Thru-Reflect-Line (TRL) calibration technique [11]. This technique requires a direct (thru) connection between the two test ports, one port reflect standard of unknown reflection at both test ports and a line standard to calibrate the VNA. The Teflon sample has been machined to fit inside the WR-90 waveguide fixture which was also used as line standard during the TRL calibration. The measurement frequency ranges from 8.2 to 12.4 GHz. The permittivity results included the dielectric constant ($\varepsilon'_r$) and the dielectric loss tangent (tan δ).

The measurement results and the uncertainty budgets for the dielectric constant and the dielectric loss tangent for the Teflon are presented here. The expanded uncertainties have been computed with a coverage factor of k=2. The behavior of $\varepsilon'_r$ and tan δ as function of frequency including the uncertainty are shown in Figs. 3 and 4.

**Fig. 3.** Dielectric constant ($\varepsilon'_r$) for the Teflon sample

**Fig. 4.** Dielectric loss tangent (tan δ) for the Teflon sample

Teflon has low dielectric constant ($\varepsilon'_r = 2.03$ [12]) and very low dielectric loss tangent in the X-band ($\tan \delta < 4 \cdot 10^{-4}$ [12], [13]). From Fig. 3, it can be seen that the measured dielectric constant results closely correspond to the known dielectric constant values for Teflon in X-band waveguide. The dielectric loss tangent shown in Fig. 4 is also lower than the typical loss tangent limit value for Teflon in X band (i.e. $\tan \delta < 4 \cdot 10^{-4}$).

The uncertainty budgets for the Teflon sample showing different uncertainty contributions for $\varepsilon'_r$ and $\tan \delta$ are shown at selected frequencies in Tables 2 and 3. The uncertainty contributions include both S-parameter uncertainties and the dimensional uncertainties. The uncertainty contributions are cable stability (Cable Stab.), connector repeatability (Conn. Rep.), drift, linearity (Lin.), noise and dimensional uncertainty (Dim. Unc.). The different uncertainty contributions have been square root summed to form the combined standard uncertainty (Std. Unc.).

**Tab. 2.** Uncertainty budget for the Teflon sample at selected frequencies for the real part of permittivity ($\varepsilon'_r$)

**Tab. 3.** Uncertainty budget for the Teflon sample at selected frequencies for the Dielectric loss tangent ($\tan \delta$)

It can be observed from Table 2 that the cable stability uncertainty has the most significant influence on $\varepsilon'_r$, while the connector repeatability, the VNA linearity and the dimensional uncertainty have significant influence on $\tan \delta$. However, compared to dimensional uncertainty, the S-parameter uncertainty has dominant contributions to the permittivity results.

## 4. Conclusion

In this paper, a fully comprehensive measurement uncertainty analysis for permittivity results has been presented. This analysis allows the optimization of the setup with regards to the measurement uncertainty by analyzing the influence of different uncertainty sources. Results are also metrologically traceable to the SI through dimensional measurements. A Teflon sample has been measured from 8.2 to 12.4 GHz using a WR-90 VNA waveguide setup. The transmission S-parameters has been used, iteratively, to compute the complex-valued permittivity. The linear propagation of uncertainties through the measurement model has been carried out. The permittivity results agree very well to the known properties of the Teflon sample in the X-band. The uncertainty budgets are also presented at different frequencies. The results highlight that the cable stability has the major influence on the dielectric constant of the Teflon sample. Therefore, one should minimize the cable movements during the VNA measurements to get accurate measurements. Dimensional measurements are also critical. For what concerns the loss tangent, the most influential contributions are related to dimensional measurements, connectors repeatability and cable stability.

# Acknowledgement

The authors are indebted with Danilo Serazio (INRIM) for providing and machining the Teflon sample. The authors are also thankful to Michael Wollensack of Federal Institute of Metrology - METAS, Switzerland for useful discussions and information.

# List of Figures

# List of Tables